\documentclass[prl,twocolumn,showpacs]{revtex4}
\usepackage{graphicx}

\begin{document}

\title{Checkerboard patterns in the t-J model}

\author{Steven R.~White}
\email{srwhite@uci.edu} 
\affiliation{Department of Physics and Astronomy, University of California\\
Irvine, CA 92697 USA}

\author{D.J.~Scalapino}
\email{djs@vulcan2.physics.ucsb.edu}
\affiliation{Department of Physics, University of California\\
Santa Barbara, CA 93106-9530 USA}

\date{\today}

\begin{abstract}

Using the density matrix renormalization group, we study the possibility
of real space checkerboard patterns arising as the ground states of
the t-J model. We find that checkerboards with a commensurate $(\pi,\pi)$
background are not low energy states and can only be stabilized with
large external potentials. However, we find that striped states with charge
density waves along the stripes can form approximate checkerboard patterns.
These states can be stabilized with a very weak external field aligning
and pinning the CDWs on different stripes.

\end{abstract}

\maketitle




Scanning tunneling microscopy (STM) studies have reported checkerboard-like
modulation patterns in the tunneling conductance of
optimally doped Bi$_2$Sr$_2$CaCu$_2$O$_{8+\delta}$ (Bi2212) near vortex 
cores \cite{Hoff02} and in nearly optimally doped Bi2212 in the absence of an
external field \cite{howald}. Checkerboard modulations have also been found in
underdoped Bi2212 \cite{Verxx,McE04} in the pseudogap region,
and in lightly doped Ca$_{2-x}$Na$_x$CuO$_2$Cl$_2$.
These modulations were found to be oriented along the
Cu-O axes. In the vortex case \cite{Hoff02}, the tunneling conductance was
modulated on a length scale of order 4.3$a$, where $a$ is the Cu-Cu
spacing.  
Some degree of one-dimensionality was observed with one Cu-O direction exhibiting
a stronger spectral intensity than the other. In their zero field studies
of Bi2212, Howald, et. al. \cite{howald} interpreted their STM measurements in
terms of a two-dimensional system of stripes with a charge modulation of $4a$.
They noted that while the modulation appeared ``almost checkerboard-like'',
the defect structure suggested that the underlying order was one-dimensional.

In experiments on underdoped Bi2212 by Vershinsin {\it et.~al}
\cite{Verxx}, the modulation length was 4.7$a \pm .2a$ and appeared only in
the normal phase at bias voltages less than a pseudogap energy of order 35 meV. 
In this case the observed conductance pattern appeared to be inherently 
two-dimensional, suggesting an absence of static stripes but leaving open the
possibility of fluctuating stripes.
In the low temperature STM studies of McElroy {\it et.~al}
\cite{McE04}, a checkerboard pattern with a spacing of order 4.5$a$ was
observed, which appeared in the underdoped nanoscale pseudogap regions at
bias voltages $\sim$ 60 meV. In Ca$_{2-x}$Na$_x$CuO$_2$Cl$_2$\cite{hanaguri},
a commensurate $4a\times 4a$ checkerboard with an additional complex incommensurate
$4a/3\times 4a/3$ internal pattern has been observed in the pseudogap regime.
In these experiments, the conductance patterns showed no
significant breaking of 90$^\circ$ rotational symmetry, suggesting a two-dimensional
checkerboard pattern as opposed to a striped pattern.

Although differing in detail,
these measurements provide evidence of an electronic locally-ordered phase
which appears when the $d_{x^2-y^2}$ superconducting phase is suppressed.
Various suggestions for this electronic phase have been put forth but at
present its nature remains unclear.  Several of these involve charge density
waves. Chen {\it et.~al} \cite{Che03,Che04} have mapped a Hubbard model
with extended Coulomb interactions onto an effective bosonic SO(5) model
and find a phase with a checkerboard density of d-wave Cooper pairs.
Anderson \cite{And04} has proposed a $4\times 4$ structure consisting of a
Wigner solid of hole pairs embedded in a sea of $d$-wave spin singlet
pairs. 
Fu {\it et.~al} \cite{Fu04} have used a Slater determinant variational ansatz
to approximate the groundstate of a generalized Hubbard model with Coulomb
and near-neighbor exchange interactions.  They find that a soliton hole
crystal phase can form with a modulation length which is $\sqrt{2}$ times
that of the d-wave pair field CDW of Chen {\it et.~al} and Anderson.  Both Chen and
Fu proposed charge density phases coexisting with a background spin state which
has commensurate $Q=(\pi, \pi)$ antiferromagnetic order. Chen {\it et.~al} 
note that in principle this is not an intrinsic
feature of their approach, but that the stability of an incommensurate
magnetic state would require extended magnetic interactions which have not
been included in their study.

Here we investigate the possibility of checkerboard order as a low energy
phase of the 2D t-J model using the density matrix renormalization 
group (DMRG)\cite{dmrg}. Our previous work using DMRG
has shown the presence of striped phases as the ground state of large
t-J clusters\cite{whitescalapino}. However, these states are sensitive to 
boundary conditions and
to additonal terms in the Hamiltonian\cite{tprime}. 
It is possible that a checkerboard phase 
could be stabilized with appropriate boundary conditions or 
small additional terms. 

It is important to specify what is meant by a checkerboard phase.
The simplest possibility, which has been the principle focus of some of the
previous theoretical work, consists of pairs of holes living primarily on
$2\times2$ placquettes, arranged in a checkerboard pattern. The
spin background in between has commensurate antiferromagnetism.
Here one imagines that an attraction between the holes has formed
the pairs, but that these pairs repel each other, forming a 
Wigner crystal-like state. In this scenario the interaction between the
pairs is something more complicated than an isotropic repulsion, so that a
checkerboard pattern results rather than a triangular
lattice of pairs. We call this phase the isotropic checkerboard phase
(ICB).

Another possibility assumes that the dominant
instability is stripe order. However, as has been observed
previously in simulations, along each stripe there is a tendency
for CDW order, associated with localized pairs in the stripe. To
make an approximate checkerboard pattern, 
one could imagine that the CDW along each stripe is pinned, and
furthermore that due to lattice or interlayer effects the CDWs on the different
stripes line up. This pattern would show two types of anisotropy:
first, the spin background would have the usual striped $\pi$-phase shifted
antiferromagnetic arrangement.  Second, the charge density
pattern would be more strongly modulated perpendicular to the
stripes. This pattern could look like an ICB phase in an STM
measurement, which would not detect the spin pattern, if the 
anisotropy were weak. We call this phase the stripe checkerboard
phase (SCB). 

Note that there is a third possibility, stemming from dynamic
stripes. Here one could imagine that the orientational order of the
stripes fluctuates, with large domains rotating by 90$^\circ$.
Since the STM measurements are made over long
time scales, they would yield a superposition of the two
orientations. To get a checkerboard pattern one would assume that
the transverse translational motion of the stripes is pinned. 
A difficulty with this approach is that the pinning potential must
be simultaneously weak enough to allow orientational fluctuations and
strong enough to pin translational motion.
As discussed below, our DMRG results for the 
t-J model in the parameter range we have studied show 
no tendency towards fluctuating orientational order.

\begin{figure}
\begin{center}
\includegraphics[ width=7cm]{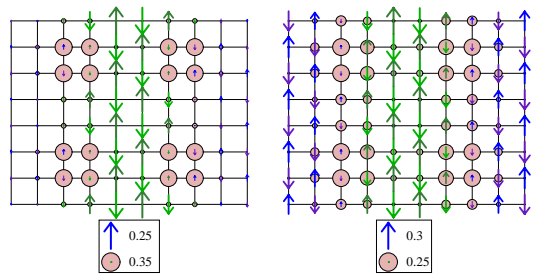}
\caption{Charge and spin configurations of a t-J model in a pinned checkerboard 
configuration.
The results show the expectation value of $<S_z>$ and $<n_h>$ for a 10x8 t-J cluster
with cylindrical boundary conditions: open in x, periodic in y. Here there are
8 holes and J/t = 0.35. Initially, a local pinning potential of -0.5 t was
placed on the 16 sites making up 4 the placquettes visible in the
left panel.  Nine sweeps were performed, keeping up to $m=1000$ states, with
the result shown in the left panel. Subsequently, the pinning
potential was turned off. 
In the right panel, we show the result after three more sweeps were 
performed, reaching m=1500 states.
} 
\label{figone}
\end{center}
\end{figure}

To study the stability of an ICB phase, we use to our advantage
a weakness of DMRG, namely that a DMRG simulation, keeping a
finite number of states, can get stuck
in a metastable ground state. For example, if on a particular
cluster a striped state oriented in the $x$-direction has
slightly lower energy than a $y$ oriented stripe state, but we
prepare the system in the $y$ oriented state, we may not see, even
after dozens of sweeps, the
transition to the $x$-orientation.  Here, we apply external fields to
stabilize an ICB state, and then remove the fields and observe
the results. In Fig. 1 we show results for a $10\times8$ t-J
cluster with $J/t=0.35$, 8 holes,  and cylindrical boundary conditions. 
The left panel shows the ICB state stabilized by a local pinning 
potential $-0.5 t (n_{i\uparrow} + n_{i\downarrow})$ 
on the 16 sites $i$ making up the four placquettes visible in the figure. The
left panel is the state after 9 DMRG sweeps, keeping up to m=1000 states. 
Note that even with the strong pinning potential, the motion of
the holes has considerably weakened the antiferromagnetism between the placquettes.
In the right panel, we show the result of the same simulation after the pinning
potential was removed and three more sweeps were performed, reaching m=1500 states.
Within one sweep of releasing the pinning, the $\pi$ phase shifts characteristic of
the stripe phase form. Also, the hole density spreads out in the y-direction
almost as quickly. We continued this run up to 17 sweeps and 3000 states. There
was very little difference between the final configuration and that shown in 
the right panel.  Results from other runs with different cluster sizes and
boundary conditions yield consistent results: the ICB phase does
not appear even as a metastable ground state. Systems prepared in
an ICB state immediately decay to an SCB ground state. 

\begin{figure}
\begin{center}
\includegraphics[ width=7cm]{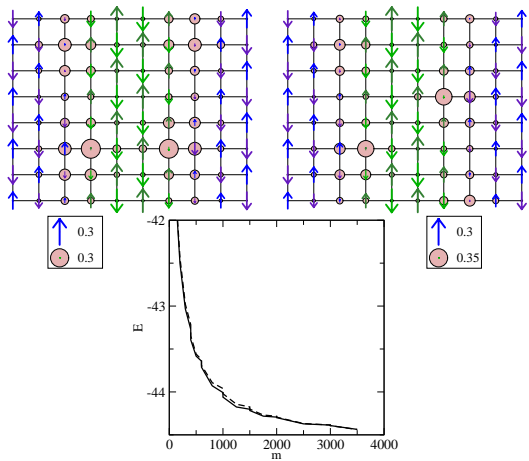}
\caption{
Charge and spin configurations of a t-J cluster with two sites pinned with a local
potential of $-0.5t$. In each panel, the sites pinned have the
large circles.  The lower plot shows the total 
energies of the systems as a function of the number of states kept per block as 
DMRG sweeps were performed on the systems. 
} 
\label{figtwo}
\end{center}
\end{figure}

However, from this we cannot conclude that the SCB state is the ground state. In
particular, the fact that the CDWs along the two stripes in the right
panel are lined up may be a residual effect of the initial
ICB state. In Fig. 2 we show results for two simulations
differing from Fig. 1 only in pinning potentials. In each case,
the initial ICB pinning potentials were not applied; instead, two
sites were given permanent pinning local potentials. In the left
panel, the two pinning sites were aligned (large circles), while
in the left, they were antialigned.  The lower plot shows the total 
energies of the systems as a function of the number of states kept per block as 
DMRG sweeps were performed on the systems. There is no significant detectable 
difference in the total energies of the two configurations. In
fact, it seems more likely that the CDWs on adjacent stripes would 
be antialigned. First, we know from previous DMRG studies that
stripes repel each other, and the simplest explanation is a Fermi
repulsion between the holes in the transverse tails of the
stripes. A CDW along a stripe would modulate this hole density,
making it look like the width of the stripe varied along its
length. In order to minimize the Fermi repulsion, we expect
antialignment of the CDWs. Second, the longer range Coulomb
repulsion terms left out of the $t$-$J$ model would certainly
favor this.

\begin{figure}
\begin{center}
\includegraphics[ width=7cm]{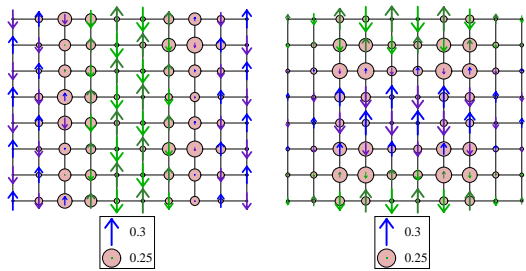}
\caption{
Charge and spin configurations with different values of the exchange coupling
linking vertical and horizontal bonds. In both cases $J_x = 0.35$. In the left
panel, $J_y = 0.37$, and in the right, $J_y = 0.38$. 
} 
\label{figthree}
\end{center}
\end{figure}

We now consider external potentials which could stabilize a checkerboard pattern. 
The ICB phase appears sufficiently unstable that it would require an 
unphysically large stabilizing potential. The SCB phase, in contrast, is stabilized
by a weak potential, which could be generated by interlayer effects or small
lattice distortions. For example, the tilt distortion in the LTT phase of 
La$_{1.6-x}$Nd$_0.4$Sr$_x$CuO is known to pin stripes. This may occur due
to Coulomb interactions, or, as discussed by Kampf, et. al\cite{kampf},
because of an anisotropic exchange interaction which arises naturally from the
lattice tilt distortion. For example, in Fig. 3, we show the result of
increasing the exchange coupling in the (spatial) $y$ direction $J_y$ slightly. 
For the isotropic case, the periodic boundary condtions in the $y$ direction
favor vertical stripes. For $J_x = 0.35$, this orientation persists up to 
about $J_y = 0.37$ (left panel). However, for $J_y = 0.38$, horizontal stripes have 
lower energy in the $10\times8$ cluster, as seen in the right panel. The transition
between the orientation seems rather sharp; there does not appear to be any
finite intermediate isotropic region, corresponding to fluctuating orientational
order.  However, one cannot draw a general conclusion from this result: other
models, including Hubbard or extended t-J models, may show different results.
Note that in the present case the
reorientation of the stripes required only a very small change in $J_x/J_y$.
Note also that an SCB pattern has appeared in the right panel. In this case, 
the open boundaries on the left and right pin and align the CDWs along each stripe.

Now consider the response of the system to a weak potential with a bond
centered spatial modulation with $Q=(0,2\pi/4a)$. Such a potential would
be expected to arise, for example, from the Coulomb interactions due
to an adjacent layer in which the stripe orientation was rotated by 90$^\circ$.
In Fig. 4 we show results from a system similar to those of Fig.
1 and 2, but with this type of applied field.  In the left panel, 
potentials of -0.05t were applied to four of the horizontal rows of sites: 
rows 2,3,6, and 7, with no potentials on the other rows.
In the right panel, we applied a potential of  $-0.1t$.
We see that a rather modest field is sufficient to stabilize the SCB pattern.

\begin{figure}
\begin{center}
\includegraphics[ width=7cm]{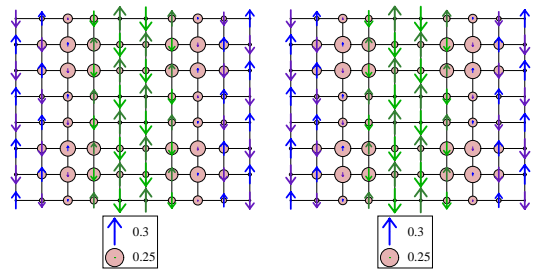}
\caption{
Charge and spin configurations with a local field in a pattern
transverse to the stripes applied. In the simulation shown in the left (right) panel, 
potentials of -0.05t (-0.1t) were applied
to four of the horizontal rows of sites: rows 2,3,6, and 7. 
} 
\label{figfour}
\end{center}
\end{figure}

In order to study the response of a stripe to a CDW inducing field with
higher precision, we
consider a two leg ladder as a model of a bond centered stripe. As is well
known, DMRG is extremely accurate on single chain and two leg ladder systems.
Here we apply a potential $\epsilon$ and measure the response $\Delta n$ 
in the central region of a $64\times2$ ladder, with a 
doping of 0.25 and $J/t=0.35$.
In a two-leg ladder\cite{cdw}, CDW and pairing correlations have competing
power law decays. Because of the power law decay, we expect a diverging
susceptibility for a CDW inducing potential. Here we are concerned
more with the size of the response to a finite potential than the
limit as the potential tends to zero. The results are shown in Fig. 5. 
We see that even a weak potential induces a strong CDW response.

\begin{figure}
\begin{center}
\includegraphics[ width=6cm]{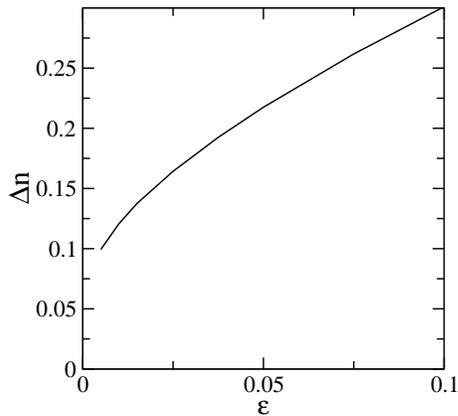}
\caption{
Charge density wave response to an applied local potential on a two leg t-J ladder,
with doping 0.25. The applied field had periodicity 4 with rungs 1 and 2 with $+\epsilon$,
rows 3 and 4 $-\epsilon$, etc. The measured response $\Delta n$ is the absolute 
difference in hole 
densities between sites 2n and 2n+1.
} 
\label{figfive}
\end{center}
\end{figure}

In summary, a real space checkerboard pattern with an antiferromagnetic spin
background does not appear to be a low energy state of the t-J model. Instead,
striped states with CDWs along the stripes can give approximate checkerboard 
patterns, but an external field, possibly arising from lattice distortions
and interplane Coulomb interations, appears
necessary to align the CDWs in each stripe.

We thank E. Demler, S.A. Kivelson, S. Sachdev, and J.M. Tranquada for useful
discussions.
We acknowledge the support of the NSF under grants DMR03-11843
and DMR02-11166.

\end{document}